\documentclass[twocolumn,showpacs,preprintnumbers,amsmath,amssymb,superscriptaddress]{revtex4}

\divide\topmargin by 3 

\usepackage{graphicx}
\usepackage{dcolumn}
\usepackage{bm}

\usepackage{amsmath}
\newcommand{\eV}{\rm eV}

\begin{document}

\title{Enhancing the spin-transfer torque through proximity of quantum
    well states}

\author{Ioannis Theodonis}
\email[E-mail: ]{ytheod@mail.ntua.gr} \affiliation{Department of
Physics, California State University, Northridge, CA 91330-8268}
\affiliation{Department of Physics, National Technical University,
GR-15773, Zografou, Athens, Greece}
\author{Alan Kalitsov}
\affiliation{Department of Physics, California State University,
Northridge, CA 91330-8268}
\author{Nicholas Kioussis}
 \affiliation{Department of Physics,
California State University, Northridge, CA 91330-8268}
\date{\today}
\begin{abstract}
{ We predict that the spin-transfer, $T_{i,||}$, and field-like,
$T_{i,\bot}$, components of the {\it local} spin torque are
dramatically enhanced in  double-barrier magnetic tunnel junctions.
The {\it spin-mixing} enhancement is due to the energetic proximity
of majority and minority quantum well states (QWS) of different
quantum numbers within the bias window. $T_{i,||}$ exhibits a
switch-on and switch-off step-like bias behavior when spin polarized
QWS enter the bias window or exit the energy band, while
$T_{i,\bot}$, changes sign between switch-on biases. The {\it net}
$T_{\bot}$ exhibits an anomalous angular behavior due to the bias
interplay of the bilinear and biquadratic effective exchange
couplings.}
\end{abstract}

\maketitle

\section{Introduction}
 Spintronics involve the exploitation of the quantum-mechanical spin degree
 of freedom to provide new functionalities beyond conventional
 electronics\cite{zutic}. One effect that has its roots on the electron's spin, is the
 torque exerted on the magnetization of a  nanometer-scale free
 ferromagnet (FM)
 by a spin-polarized current, originating from a
 preceding non-collinear pinned FM\cite{Slonczewski1,Slonczewski2,berger}. This torque can
be decomposed into a field-like and a spin-transfer
component\cite{Zhang}, both orthogonal to the magnetic moment of the
free FM, but with different influence on its
dynamics\cite{stilesnew}. Recent ferromagnetic resonance
experiments, provide a useful tool to study the role of each
component\cite{tulapurkar,sankey}. At sufficiently high current
densities, the spin-transfer torque leads to current-induced
magnetization switching (CIMS)\cite{Fuchs,Urazhdin,huai}. Reduction
of the high critical current for CIMS is necessary for spin-transfer
controlled magnetic memories\cite{sony}.


\begin{figure}[tt]
\includegraphics[width=8.5cm]{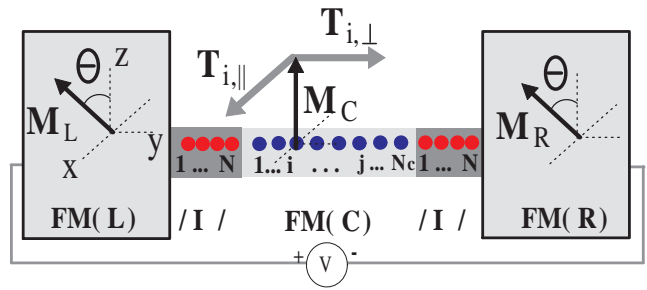}
\caption{(Color Online) Schematic of the DBMTJ consisting of a FM
central wire of $N_C$ atomic sites, connected to left and right FM
leads through the tunneling barriers $I$ of $N$ sites.
The spin-transfer (field-like), $T_{i,||}$ ($T_{i,\bot}$) components
of the torque lies in the -x (y) directions.}\label{fig-0}
\end{figure}

Double-barrier magnetic tunnel junctions (DBMTJ) consist of a
central metallic layer between two insulating barriers and two FM
electrodes. The tunneling magnetoresistance (TMR) can be
dramatically enhanced in {\it collinear} DBMTJ by the presence of
quantum well states (QWS) under appropriate resonant
conditions\cite{Petukhov,pantelidis,zeng,nozaki}. Recently, the
discrete energy spectrum of FM nanoparticles has been shown to
enhance spin accumulation and to control the bias dependence of the
TMR\cite{Yakushiji}. While the physics in {\it collinear} DBMTJ has
been studied extensively\cite{Petukhov,pantelidis,zeng,nozaki}, the
effect of {\it spin-polarized} QWS (SPQWS) on the spin-torque in
{\it non-collinear} DBMTJ remains an unexplored area thus far.

The objective of this work is to present for the first time a study
of the effect of SPQWS on both the {\it spin-transfer} $T_{||}$ and
{\it field-like} $T_{\bot}$ components of the spin torque under
external bias. The calculations are based on the tight-binding
method and the non-equilibrium Keldysh formalism. We predict that
both components of the {\it local} spin torque can be dramatically
enhanced when majority and minority QWS energies of different
quantum numbers are in close proximity and lie within the bias
window. It should be emphasized that the local-spin-torque
enhancement is not associated with an enhancement of the
corresponding spin-polarized currents. The low-temperature bias
dependence of the local spin-transfer torque,$T_{i,||}$, exhibits a
switch-on and switch-off step-like behavior when the SPQWS enter the
bias window or exit the energy band, respectively, similar to that
of the spin-polarized currents. On the other hand, $T_{i,\bot}$
changes sign between the majority and minority switch-on values of
bias. We demonstrate that the bias behavior of $T_{i,||}$ and
$T_{i,\bot}$ can be derived analytically using a single-site central
FM region. The {\it net} $T_{\bot}$, pertinent to the
non-equilibrium interlayer exchange coupling, $E_{XC}$, exhibits an
anomalous angular behavior due to the interplay of the bilinear and
biquadratic effective exchange couplings which have different bias
behavior.

The structure of this paper is as follows. In Sec. II, we introduce
the basic model  and outline the computational approach of the
non-equilibrium spin torque. In Sec. III, we present and discuss the
results of the  calculations. Finally, Sec. IV includes a brief
statement of the conclusions.
\section{Method - Formalism}
Fig.\ref{fig-0} shows the one-dimensional FM/I/FM/I/FM DBMTJ system,
consisting of a central (C) FM nanowire  containing $N_C$ atomic
sites (AS) connected to the left (L) and right (R) FM electrodes
through two thin symmetric non-magnetic tunneling barrier nanowires
$I$ of $N=2$ AS. The magnetization of the central FM, ${\bf M}_C$,
is along the $z$ axis of the coordinate system shown in Fig.
\ref{fig-0}. The magnetization of the FM leads  ${\bf M}_{L(R)}$
lies in the $x-z$ plane, i.e. it is rotated by angle $\theta$ around
the wire axis $y$.
The Hamiltonian of the system is
\begin{equation}
H=H_L+H_{M}+H_R+H_{L,M}+H_{R,M} + hc ,
\end{equation}
where $H_M = H_{I}+H_C+H_{I,C}+ hc$ is the Hamiltonian of the middle
(M) multi-layer $I$/FM/$I$ region, and $H_{L(R),M}, H_{I,C}$ are the
coupling Hamiltonians at the L(R)/M and I/C interfaces,
respectively. The Hamiltonian for each FM region $H_\alpha$,
$\alpha$= L, R, and C, is described by a one-dimensional
single-orbital tight-binding model neglecting the in-plane
k-dependence, which includes a nearest-neighbor (NN)
spin-independent hopping term, $t_{\alpha}$, and a spin-dependent
on-site energy term, $\varepsilon_{\alpha}^{\sigma}$, i.e.,
\begin{equation}
H_\alpha = \sum_{\sigma,i}\varepsilon_\alpha^\sigma c_i^\dagger
c_i + \sum_i t_\alpha c_i^\dagger c_{i+1} +hc.
\end{equation}
The Hamiltonian, $H_{I}$, for the barriers is identical to
$H_\alpha$, but where one replaces the hopping term $t_\alpha$ with
$t_I$ and the spin-dependent on-site energy
$\varepsilon_\alpha^\sigma$ with the spin-independent
$\varepsilon_I$. The coupling Hamiltonian of the $FM/I$ interface is
$H_{\alpha,I}=t_{\alpha/I} c_\alpha^\dagger c_I$. The
exchange-splitting, $\Delta_{\alpha} = \varepsilon_{\alpha}^\uparrow
- \varepsilon_{\alpha}^\downarrow$, is identical in all FMs,
$\Delta_{\rm I}=0$, $\varepsilon_{\alpha}^\uparrow - E_F =0.318 ~\eV
$, $\varepsilon_{\alpha}^\downarrow -E_F =0.736~\eV $, and
$\varepsilon_{\rm I} -E_F =6.5~\eV$, where $E_F$ is the Fermi
energy. The $t_{\alpha} = 0.4~\eV$ in all FMs, $t_{I}=1~\eV$ in both
insulators, and $t_{\alpha/I} = 0.5~\eV$ in all FM/I interfaces,
consistent with the {\it ab initio} values for 1D Co FM
nanowires\cite{hong}.

We extend Datta's formalism~\cite{datta} to the case of
non-collinear systems, where the scalar Green functions are replaced
with 2x2 matrices in spin space. For this purpose, $H_\alpha$ can be
expressed in the form, $H_\alpha = \bar H_{\alpha} +  \delta
H_{\alpha}$, where
\begin{equation}
\begin{array} {ccc}
\bar H_{\alpha} = \frac{1}{2}(\varepsilon_\alpha^\uparrow +
\varepsilon_\alpha^\downarrow)+ t_\alpha,
\end{array}
\end{equation}
describes the spin-average part of $H_\alpha$ and
\begin{equation}
\begin{array} {ccc}
\delta H_{\alpha} = \frac{1}{2}(\varepsilon_\alpha^ \uparrow  -
\varepsilon_\alpha^ \downarrow )
\end{array}
\end{equation}
is the spin-splitting part of $H_\alpha$. The one-electron
Schr$\ddot{o}$dinger equation for the retarded Green function,
$g_{pq}^{\sigma,\sigma'}$, in each isolated semi-infinite
ferromagnetic lead becomes
\begin{equation}
\begin{array} {ccc} \sum_{p_1} \left [ (E
\delta_{pp_1}-\stackrel-{H}_{pp_{1}})I-\delta
H_{pp_{1}}
\left
( \begin{array} {ccc} cos \theta & sin \theta \\ sin \theta & -cos
\theta \\ \end{array} \right ) \right ]  \times
   \\ \left ( \begin{array} {ccc} g_{p_{1}q}^{\uparrow \uparrow} &
   g_{p_{1}q}^{\uparrow
   \downarrow}
   \\
g_{p_{1}q}^{\downarrow \uparrow} & g_{p_{1}q}^{\downarrow
\downarrow} \\ \end{array} \right ) = \delta_{pq} I,   \\
\end{array}\label{leadsHamiltonian}
\end{equation}
where $I$ is a 2x2 unit matrix. Following Datta~\cite{datta} we find
that the retarded Green function of the middle region M is
\begin{equation}
\hat G_M= [E\hat I-\hat H_M-\hat \Sigma_L-\hat \Sigma_R]^{-1},
\end{equation}
where $E$ is the one-electron electron energy, $\hat H_M$ and
$\hat \Sigma_{L(R)}$ are the $(2N_M\times2N_M)$ Hamiltonian and
self-energy matrices, respectively, and $N_M= 2N + N_C$ is the
number of atomic sites in the middle region. The only non-zero
elements of $\hat \Sigma_{L(R)}$ are the $(2\times2)$ self-energy
matrices at the interfacial sites
\begin{equation}
\tilde{\Sigma}_{L(R)}(\theta)= t_{\alpha/I}^2 \tilde{
g}_{L(R)}(\theta),
\end{equation}
where $\tilde{g}_{L(R)}(\theta)$ are the retarded surface
$2\times2$ Green's function matrices of the isolated L(R) lead,
determined from Eq. \ref{leadsHamiltonian}. The non-equilibrium
Green's functions can be determined by solving the kinetic
equation~\cite{datta}
\begin{equation}
\hat G_M^<=i\hat G_M\hat\Sigma^<\hat G_M^\dagger,
\end{equation}
where $\hat\Sigma^<=f_L(\hat\Sigma_L^\dagger-\hat\Sigma_L)+f_R
(\hat\Sigma_R^\dagger-\hat\Sigma_R)$, is the non-equilibrium
self-energy matrix and $f_{L(R)}$ are the Fermi-Dirac distribution
functions of the L(R) leads.

The local spin-transfer torque ${\bf T}_i$ exerted on the local
moment at site $i$ in the central FM region is\cite{theodonis}
\begin{equation}
{\bf T}_i \equiv
-\nabla\cdot\mbox{\boldmath$I$}^{(s)}=\mbox{\boldmath$I$}^{(s)}_{i-1,i}-
\mbox{\boldmath$I$}^{(s)}_{i,i+1},
\end{equation}
 where
\begin{equation}
\mbox{\boldmath$I$}^{(s)}_{i,i\pm 1} = \frac{t_C}{4\pi} \int
Tr_\sigma\left [ (\tilde{G}^<_{i,i\pm 1} - \tilde{G}^<_{i\pm
1,i})\mbox{\boldmath$\sigma$} \right ] dE \label{spincurrent},
\end{equation}
is the spin current between NN sites\cite{theodonis}, and
$\mbox{\boldmath$\sigma$}=(\sigma_x,\sigma_y,\sigma_z)$ is a vector
of the Pauli matrices. The field-like, $T_{i,\bot}$,  and
spin-transfer, $T_{i,\|}$, components of the local spin torque,
shown in Fig. \ref{fig-0}, are along the
$\hat{M}_{C}\times(\hat{M}_{L(R)}\times\hat{M}_{C})$ and
$\hat{M}_C\times\hat{M}_{L(R)}$ directions, respectively.
Here, $\hat{M}_C$ and
$\hat{M}_{L(R)}$ are unit vectors along the magnetization of the
free C and pinned L(R) FM regions, respectively.

\section{Results - Discussion}

\begin{figure}[tt]
\includegraphics[width=8.5cm]{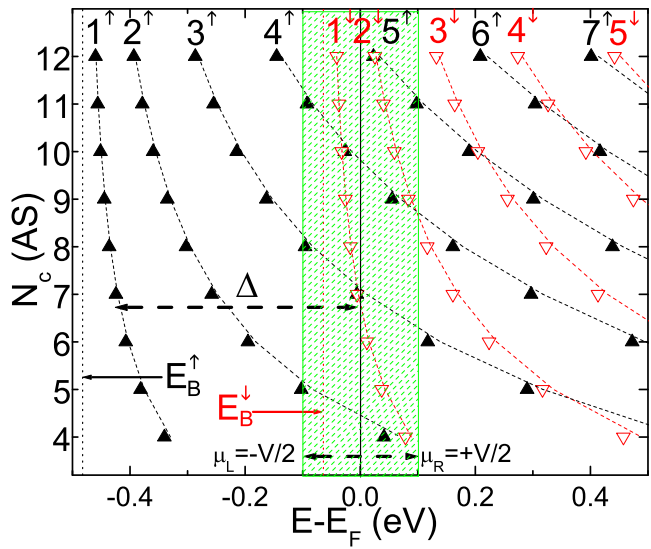}
\caption{(Color online) SPQWS energy positions $E^{n^\sigma}$
as a function of the number of atomic sites, $N_c$, in the central
FM region for  zero bias and $\theta=0$. The bottom of the majority
(minority) conduction band of the leads is denoted by
$E^{\uparrow(\downarrow)}_B$ and $\Delta$ is the exchange splitting,
denoted by the dashed horizontal arrow. At finite bias $V$, the
chemical potentials of the L,R leads are shifted by $eV=\mu_R-\mu_L$
around the Fermi energy.\label{fig-2} }
\end{figure}

The majority-(full triangles) and minority-(open triangles) QWS
 energies relative to the Fermi energy, $E^{n^\sigma}$, as a function of
the thickness, $N_C$, of the central FM wire are shown in
Fig.~\ref{fig-2} for QWS between -0.5 eV and 0.5 eV. The numbers
next to each series of data points, indicate the quantum number,
$n^\sigma = 1^\sigma, 2^\sigma,\ldots, N_c^\sigma$, of the SPQWS.
The dashed curves denote the SPQWS energies,
$E^{n^\sigma}_{0}=\varepsilon^\sigma+2t\cos(n^\sigma\pi/(N_C+1))$
of the isolated central FM wire. The coupling of the central
region to the FM leads results in a shift and a broadening of the
SPQWS energies. The bottom of the majority (minority) conduction
band of the leads, at zero bias, is indicated by
$E^{\uparrow(\downarrow)}_B$. Note, that for $N_c=7$AS the
$n^{\uparrow}=3^{\uparrow}$ majority- and
$n^\downarrow=1^\downarrow$ minority QWS are in very close
proximity and they are very close to the Fermi energy. Under
applied bias $V$, only the SPQWS with energies $E^{n^\sigma}$ that
lie within the bias window from $\mu_L=-\frac{eV}{2}$ to
$\mu_R=+\frac{eV}{2}$, denoted by the shaded area in Fig.
\ref{fig-2}, contribute to the resonant tunneling.
\begin{figure}[tt]
\includegraphics[width=8.5cm]{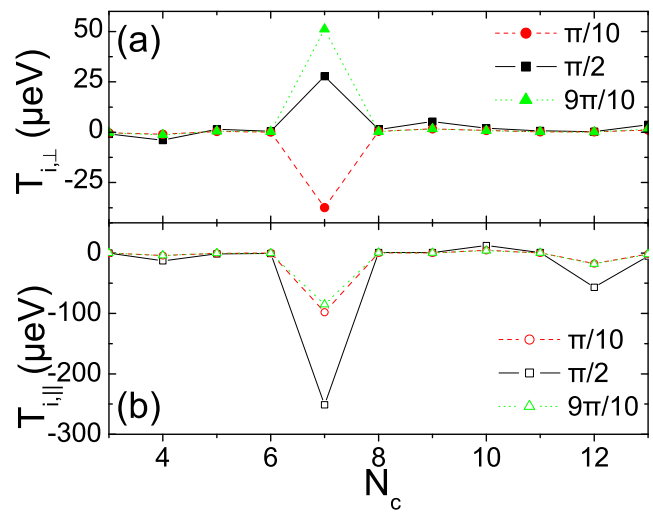}
\caption{(Color online) Low-temperature (T=5 K) and
low-bias (V = 0.1 V) (a) perpendicular, $T_{i,\bot}$, and (b)
parallel, $T_{i,||}$, components of the local spin torque
versus the number of atomic sites, $N_c$, of the central
FM region. Both components of the local spin torques
are calculated on the first site in the central FM region next to
the left FM/I interface for three values of the angle $\theta$ of $\pi/10$, $\pi/2$, and
$9\pi/10$, respectively.}\label{newfig-3}
\end{figure}

In Figs. \ref {newfig-3}a and \ref{newfig-3}b we show the
perpendicular, $T_{i,\bot}$, and parallel, $T_{i,||}$, components of
the local spin torque on the first site ($i=1$) in the central FM
region next to the left FM/I interface, as a function of the
thickness $N_c$ of the central FM region. Both local spin torque
components are calculated at T=5K and V=0.1V, for the almost
parallel $\theta=\pi/10$, perpendicular $\theta=\pi/2$ and almost
antiparallel $\theta=9\pi/10$ configurations. We find that
$T_{i,\bot}$ and $T_{i,||}$ are strongly enhanced for $N_c=7$ AS by
about one and two orders of magnitude, respectively. The local
spin-torque enhancement persists even for very small angular
deviations from the parallel(circles) and antiparallel(triangles)
configurations. Interestingly, $T_{i,\bot}$ for $N_c = 7$ changes
sign with increasing $\theta$, in contrast to $T_{i,||}$.

\begin{figure}[tt]
\includegraphics[width=8.5cm]{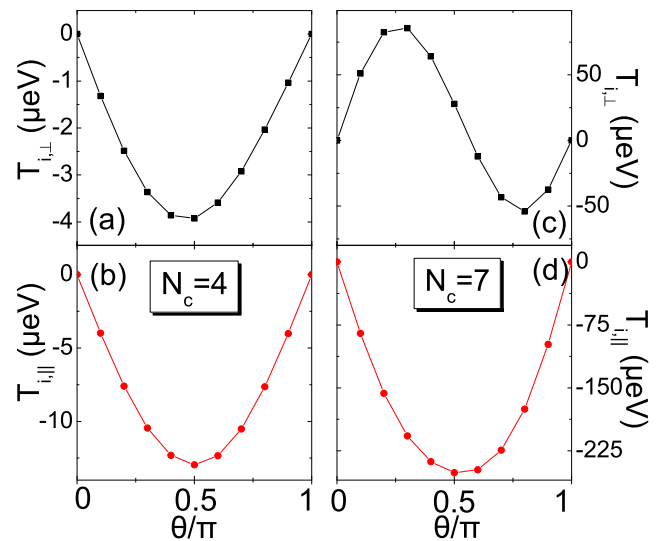}
\caption{(Color online) Angular dependence of $T_{i,\bot}$ (black
squares) and $T_{i,||}$ (red circles) for $N_c=4$AS in panels (a)
and (b), and for $N_c=7$AS in panels (c) and (d), respectively. The
local spin torque is evaluated on the first
 site in the central FM region next to the left FM/I interface
at $T=5 K$ and $V = 0.1V$.}\label{newfig-4}
\end{figure}
\begin{figure}[tt]
\includegraphics[width=8.5cm]{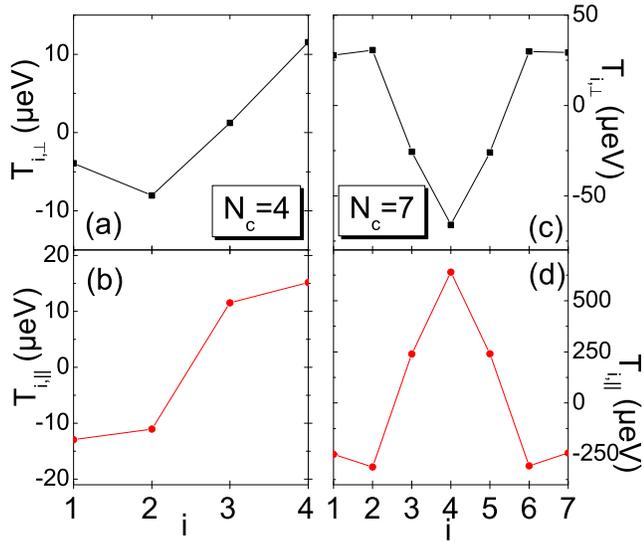}
\caption{(Color online) Perpendicular, $T_{i,\bot}$ (black squares,
and parallel, $T_{i,||}$ (red circles), components of spin torque as
a function of site $i$ in the central FM region,  for $N_c=4$AS in
panels (a) and (b) and for $N_c=7$AS in panels (c),(d),
respectively. The local components are calculated at T=5K, 0.1 V
bias, and $\theta=\pi/2$.}\label{newfig-5}
\end{figure}

In Figs. \ref{newfig-4}a,\ref{newfig-4}c and
\ref{newfig-4}b,\ref{newfig-4}d we display the angular dependence of
the perpendicular, $T_{i,\bot}$, (black squares) and parallel
$T_{i,||}$ (red circles) components of the {\it local} spin torque
for $N_c=4$AS and $N_c=7$AS, respectively. The spin torques are
evaluated on the first site in the central FM region next to the
left FM/I interface at T=5K and V=0.1V. In both cases (Figs.
\ref{newfig-4}b and \ref{newfig-4}d) $T_{i,||}$, exhibits a
sinusoidal angular dependence similar to that in a single
MTJ\cite{Slonczewski2}. On the other hand, when the enhancement
conditions are fulfilled for $N_c=7$ AS (Fig. \ref{newfig-4}c),
$T_{i,\bot}$ exhibits a non-sinusoidal angular dependence changing
sign in the $[0,\pi]$ interval.

In Figs. \ref{newfig-5}(a),(c) and \ref{newfig-5}(b),(d) we display
the perpendicular, $T_{i,\bot}$, (black squares) and parallel
$T_{i,||}$ (red circles) components of the {\it local} spin torque
as a function of site $i$ in the central FM, for $N_c=4$AS and
$N_c=7$AS, respectively. The spin torques are evaluated at T=5K,
V=0.1V, and $\theta=\pi/2$. Interestingly both $T_{i,\bot}$ and
$T_{i,||}$ oscillate around zero as function of atomic site $i$ due
to the electron precession in the central FM. The number of nodes
increases  as the width of the FM quantum well increases. The
enhancement of $T_{i,\bot}$ and $T_{i,||}$ for $N_c=7$ AS holds for
all sites $i$.


\begin{figure}[tt]
\includegraphics[width=8.5cm]{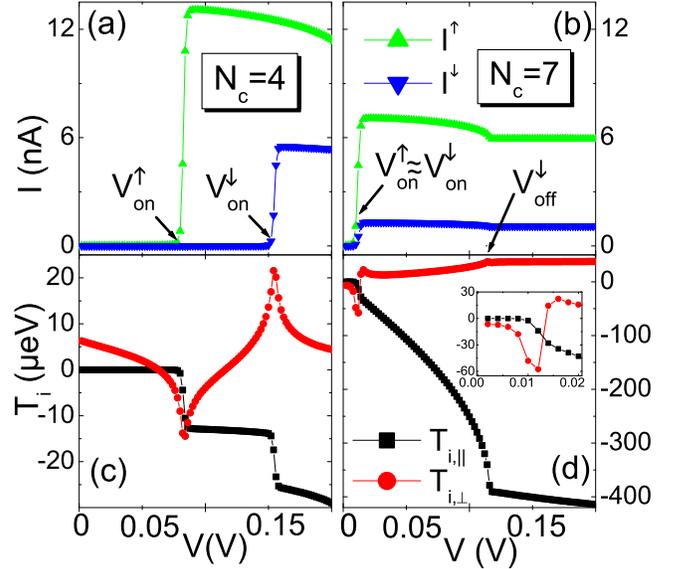}
\caption{(Color online) Low-temperature (T=5K) bias dependence of
the majority (green symbols) and minority (blue symbols) currents in
the central FM for (a) $N_c=4$ AS and (b) $N_c=7$AS, respectively,
and for $\theta=\pi/2$. Bias dependence of $T_{i,||}$ (black
symbols) and  $T_{i,\bot}$ (red symbols)
 on the first site in the
central FM, for (c) $N_c=4$ AS  and (d) $N_c=7$. The bias
$V_{on(off)}^{\sigma}$ denote the switch-on (-off)
bias.}\label{fig-3}
\end{figure}

In Figs. \ref{fig-3}a and \ref{fig-3}b, we display the
low-temperature ($T = 5K$) bias dependence of the spin-polarized
currents, $I^{\uparrow}$ and $I^{\downarrow}$, for $ N_c=4$AS and
$N_c=7$AS, respectively, and for $\theta = \frac{\pi}{2}$. The
spin-polarized currents switch on at
$V_{on}^\sigma=2|E^{n_{^\sigma}}-E_F|$, when the $n_{\sigma}$ QWS
enters the bias energy window. For $N_c=7$AS, the switch on for both
spin-polarized currents occur at about the same bias,
$V_{on}^\uparrow \approx V_{on}^\downarrow$, due to the fact that
$|E^{1^\downarrow}-E_F| \approx |E^{3^\uparrow}-E_F|$ in
Fig.~\ref{fig-2}. Both currents decrease with increasing bias
because the density of states of the minority band in the leads at
$E^{n^{\sigma}}$ decreases. At the critical bias,
$V_{off}^{\downarrow} = 2(E^{n^\sigma} + |E^{\downarrow}_B|)$, the
QWS energies, $E^{1^{\downarrow}}$ and $E^{3^{\uparrow}}$ are
shifted below the bottom of the minority band of the lead, and hence
the minority contribution to the spin-polarized currents is switched
off.

In Figs. \ref{fig-3}c and \ref{fig-3}d we display the
low-temperature ($T = 5K$) bias behavior of the parallel,
$T_{i,||}$, (black squares) and perpendicular $T_{i,\bot}$ (red
circles) components of the {\it local} spin torque on the first site
in the central FM, for $\theta = \frac{\pi}{2}$ and for $ N_c=4$AS
and $N_c=7$AS, respectively. The local spin-transfer component,
$T_{i,||}$, exhibits a switch on bias behavior at $V_{on}^\uparrow$
and at $V_{on}^\downarrow$, similar to that of the spin polarized
currents in \ref{fig-3}a. On the other hand, $T_{i,\bot}$, which is
non-zero for zero bias, displays a non-monotonic bias dependence,
changing sign between $V_{on}^\uparrow$ and $V_{on}^\downarrow$,
similar to that of the exchange field in quantum dots connected to
FM leads\cite{braun}. It is important to note that both $T_{i,||}$
and $T_{i,\bot}$ are strongly enhanced for $N_c=7$AS in
Fig.~\ref{fig-3}d, even though the corresponding spin-polarized
currents are {\it smaller} than those for $N_c=4$ AS. Thus, the
enhancement of the local spin torque is not associated with a
corresponding enhancement of the spin-polarized currents. This
result clearly demonstrates that the underlying mechanism that
controls the {\it local} spin-transfer torque {\it enhancement} is
the close proximity of the majority and minority QWS energies of
different quantum number, $E^{n^{\uparrow}}\approx
E^{n'^{\downarrow}}$, within the bias energy window. This in turn
enhances the spin mixing $\sigma \leftrightarrow \bar\sigma$ in the
central FM, when electrons tunnel resonantly through the SPQWS. The
enhancement of the local spin-transfer torque is independent of the
parity of the QWS wavefunctions. This spin-transfer torque
enhancement may have technological applications since the CIMS may
be facilitated under such conditions.
\begin{figure}[tt]
\includegraphics[width=8.5cm]{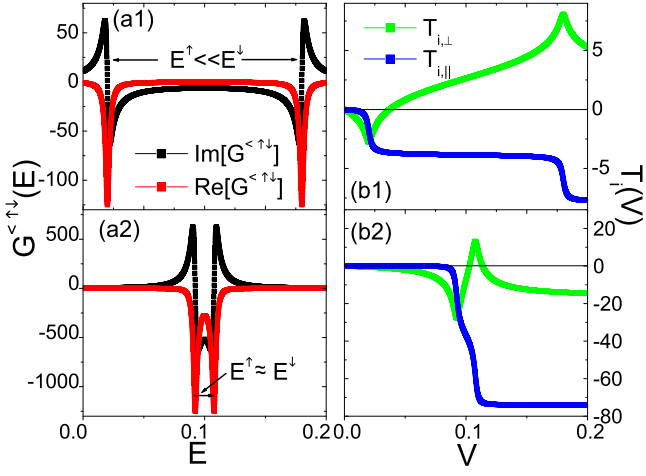}
\caption{(Color Online)(a1),(a2) Energy dependence of the real (red
squares) and imaginary (black squares) parts of
$G^{<\uparrow\downarrow}$ for ($E^\uparrow << E^\uparrow$) and
($E^\uparrow \approx E^\uparrow$), respectively;(b1) and (b2) bias
dependence of $T_{i,||}$ (blue squares) and of $T_{i,\bot}$ (green
squares) corresponding to (a1) and (a2), respectively.}\label{fig-4}
\end{figure}

In order to elucidate the role of the  SPQWS on the enhancement and
bias dependence of the local spin torque, we have used a simple
model of a central FM region consisting of a single-site coupled
weakly to the leads. We have shown that to leading order in the
coupling parameter, $t_C$, the on-site spin off-diagonal
non-equilibrium Greens function is
\begin{equation}
G^{<\uparrow\downarrow}\approx i \sum_{\alpha=L,R} f_{\alpha}
G_{r}^{\uparrow\uparrow}\Sigma_{\alpha}^{\uparrow\downarrow}
G_{a}^{\downarrow\downarrow} \label{simplegmp}.
\end{equation}
Here, $\Sigma_{\alpha}^{\uparrow\downarrow}=t_{C}^2\pi
[N^\uparrow_{\alpha}- N^\downarrow_{\alpha}]sin(\theta)$ is the
self-energy due to the leads, $N^{\uparrow(\downarrow)}_{\alpha}$ is
the surface density of states of the L(R) leads, which for
simplicity are taken to be energy independent. In the weak coupling
regime, the retarded (advanced) Greens functions of the coupled
system, $G^{\sigma\sigma}_{r(a)}$, can be be approximated with those
of the uncoupled system $g^{\sigma\sigma}_{r(a)}$, i.e.
$G_{r(a)}^{\sigma\sigma}\approx g^{\sigma\sigma}_{r(a)}=
[E-E^{\sigma}\pm i\eta]^{-1}$, where $\eta$ is taken to be
spin-independent. Therefore, the spin-transfer torque components,
determined by the non-equilibrium on-site
magnetization~\cite{kalitsov}, are
\begin{equation}
T_{i,||}\propto \int_{-\infty}^{eV} \frac{\eta
(E^\uparrow-E^\downarrow)}{((E-E^\uparrow)^2+
\eta^2)((E-E^\downarrow)^2+\eta^2)}dE, \label{simplegmpRe2}
\end{equation}
and
\begin{equation}
T_{i,\bot}\propto \int_{-\infty}^{eV}
\frac{(E-E^\uparrow)(E-E^\downarrow)+\eta^2}{((E-E^\uparrow)^2+
\eta^2)((E-E^\downarrow)^2+\eta^2)}dE \label{simplegmpIm2}.
\end{equation}

The energy dependence of the real (red squares) and imaginary (black
squares) parts of $G^{<\uparrow\downarrow}$ for ($E^\uparrow <<
E^\uparrow$) and ($E^\uparrow \approx E^\uparrow$) are shown in
panels (a1) and (a2) of Fig.~\ref{fig-4}, respectively. The
corresponding bias dependence of $T_{i,||}$ (blue squares) and of
$T_{i,\bot}$ (green squares) are shown in panels (b1) and (b2) of
Fig.~\ref{fig-4}, respectively. One can clearly see that the overall
bias dependence of this simple model reproduces qualitatively that
displayed in Figs.\ref{fig-3}c and \ref{fig-3}d. More specifically,
the switch-on behavior for $T_{i,||}$ and the sign change of
$T_{i,\bot}$ with bias are associated with the relative position of
the majority and minority QWS energies which lie within the bias
energy window. When $E^\uparrow \approx E^\downarrow$, both
components of the spin-transfer torque are dramatically enhanced, as
shown in Fig.~\ref{fig-4}(a2), due to the presence of higher-order
poles in Eqs. (\ref {simplegmpRe2}) and (\ref{simplegmpIm2}).

\begin{figure}[tt]
\includegraphics[width=8.5cm]{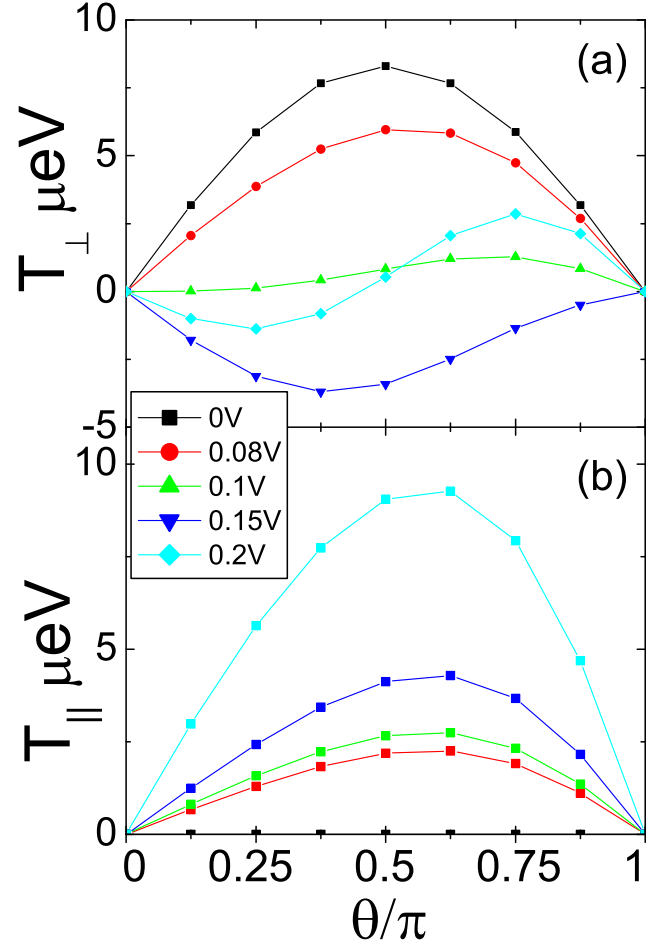}
\caption{(Color Online)(a) Angular dependence of the {\it net} (a)
field-like torque and (b) spin-transfer torque, for $N_c=4$ and
various values of bias.}\label{newfigure8}
\end{figure}

\begin{figure}[tt]
\includegraphics[width=8.5cm]{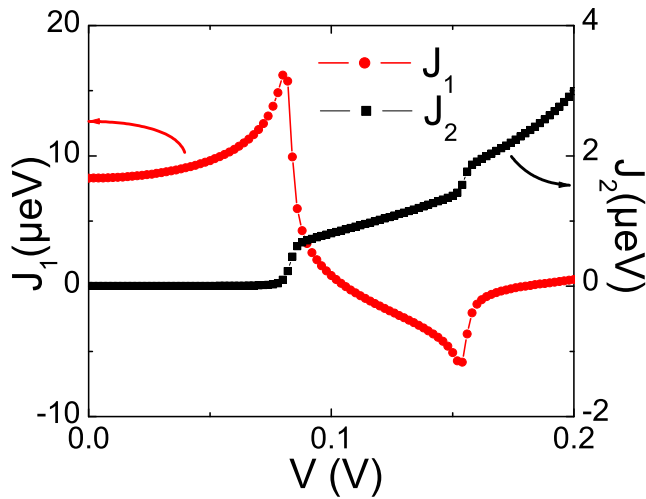}
\caption{(Color Online) Bias dependence of the non-equilibrium
effective bilinear, $J_1$, and biquadratic, $J_2$, interlayer
exchange couplings for $N_c=4$.} \label{newfigure9}
\end{figure}

 The {\it net} spin torque components $T_{||(\bot)} =
\sum_{i \varepsilon C}T_{i,||(\bot)}$, are not as strongly
enhanced as the local torques, $T_{i,||}$ and $T_{i,\bot}$, which
oscillate as a function of site $i$, due to the electron
precession in the central FM (Fig.\ref{newfig-5}). The angular
dependence of the {\it net} field-like torque, $T_{\bot}$, and the
spin-transfer torque $T_{||}$, is shown in Fig. \ref{newfigure8}a
and \ref{newfigure8}b, respectively, for $N_c=4$ AS and for
various bias values. For zero bias $T_{\bot}$ displays sinusoidal
behavior, which however, can change dramatically upon increasing
the bias. The {\it net} field-like torque can be
expressed\cite{Slonczewski1,Albuquerque} as
\begin{equation}
T_{\bot}= -\partial E_{XC}(\theta)/\partial\theta,
\end{equation}
where
\begin{equation}
E_{XC}(\theta)=-J_1 cos(\theta) - J_2 cos^2(\theta) + \ldots,
\end{equation}
 is the {\it effective} exchange coupling energy
\cite{SloncezwskiBIquad}, between ${\bf M}_C$ and ${\bf M}_{L(R)}$.
The out of equilibrium interlayer exchange coupling has terms
related to the spin-polarized tunnel current that can dominate and
alter the coupling behavior under certain bias
conditions\cite{heide}. Here, $J_1$ and $J_2$ are the {\it non
equilibrium} bilinear and biquadratic effective exchange couplings,
respectively, which are determined by fitting the angular dependence
of $T_{\bot}(\theta)$ in Fig.~\ref{newfigure8} to the above
expression for various biases. In Fig.~\ref{newfigure9} we show the
bias dependence of $J_1$ and $J_2$ for $N_c=4$ AS. It is important
to note that $J_1$ (red circles) reverses its sign with bias,
similar to $T_{i,\bot}(V)$ in Fig.~\ref{fig-3}c. On the other hand,
the bias dependence of $J_2$ (black squares) exhibits the switch on
bias behavior found for $T_{i,||}(V)$. Hence, there is a range of
bias where $J_2>J_1$, favoring perpendicular alignment of ${\bf
M}_C$ and ${\bf M}_{L(R)}$\cite{SloncezwskiBIquad}. In contrast, the
angular dependence of the {\it net }$T_{||}$ exhibits a skewed
sinusoidal behavior (\ref{newfigure8}b) for any bias, similar to
that found in spin valves.

\section{Conclusion}
In summary, we demonstrate that the local spin-transfer torque can
be dramatically enhanced, when the majority and minority QWS of
different quantum number are in close energetic proximity and lie
within the bias energy window. This enhancement may in turn lead to
reduction of the critical current necessary for CIMS in magnetic
memories. The spin-torque enhancement criterion may be achieved by
controlling the SPQWS through an external magnetic field or
spin-dependent barriers. The SPQWS tune selectively the bias
dependence of the spin-transfer and field-like components of the
{\it local} and the {\it net} spin torque. This results to
 an anomalous angular behavior of $T_{\bot}$
due to the bias interplay of the bilinear and biquadratic effective
exchange couplings. Future work will be aimed to include the results
for the local spin torques of these calculations as an input into
the Landau-Lifshitz-Gilbert equation, to calculate the critical
current for the CIMS.
\section{Acknowledgements}
The research was supported by NSF-PREM grant DMR-0611562, US Army
grant W911NF-04-1-0058, and NSF-KITP grant PHY99-07949.

\end{document}